\documentclass[aps,prd,floatfix,nofootinbib,superscriptaddress,twocolumn,10pt]{revtex4}
\usepackage{graphicx}
\usepackage{bm}
\usepackage{subfigure}
\usepackage{amssymb}
\usepackage{color,soul}
\usepackage{graphicx}
\usepackage{multirow}
\usepackage{amsmath}
\usepackage[colorlinks,citecolor=black,urlcolor=black,linkcolor=black]{hyperref}
\DeclareUnicodeCharacter{0301}{\'{e}}
\graphicspath{{figure/}}
\DeclareGraphicsExtensions{.pdf,.png,.jpg}
\setulcolor{red}

\newcommand{\COLOSSUS}[1]{\href{https://bdiemer.bitbucket.io/colossus/}{\bf COLOSSUS}}

\newcommand{\CLASS}[1]{\href{https://lesgourg.github.io/class_public/class.html}{\bf CLASS}}

\newcommand{\beq}{\begin{equation}}
\newcommand{\eeq}{\end{equation}}
\newcommand{\beqa}{\begin{eqnarray}}
\newcommand{\eeqa}{\end{eqnarray}}

\begin{document}
\title{
Mirror QCD phase transition as the origin of the nanohertz Stochastic Gravitational-Wave Background
}

\author{Lei Zu }

\affiliation{Key Laboratory of Dark Matter and Space Astronomy, Purple Mountain Observatory, Chinese Academy of Sciences, Nanjing 210023, China}

\author{Chi Zhang}

\affiliation{Key Laboratory of Dark Matter and Space Astronomy, Purple Mountain Observatory, Chinese Academy of Sciences, Nanjing 210023, China}
\affiliation{School of Astronomy and Space Science, University of Science and Technology of China, Hefei 230026, China}

\author{Yao-Yu Li}

\affiliation{Key Laboratory of Dark Matter and Space Astronomy, Purple Mountain Observatory, Chinese Academy of Sciences, Nanjing 210023, China}
\affiliation{School of Astronomy and Space Science, University of Science and Technology of China, Hefei 230026, China}

\author{Yuchao Gu }

\affiliation{Key Laboratory of Dark Matter and Space Astronomy, Purple Mountain Observatory, Chinese Academy of Sciences, Nanjing 210023, China}

\author{Yue-Lin Sming Tsai\footnote{Corresponding Author: smingtsai@pmo.ac.cn}}

\affiliation{Key Laboratory of Dark Matter and Space Astronomy, Purple Mountain Observatory, Chinese Academy of Sciences, Nanjing 210023, China}
\affiliation{School of Astronomy and Space Science, University of Science and Technology of China, Hefei 230026, China}

\author{Yi-Zhong Fan\footnote{Corresponding Author: yzfan@pmo.ac.cn}}

\affiliation{Key Laboratory of Dark Matter and Space Astronomy, Purple Mountain Observatory, Chinese Academy of Sciences, Nanjing 210023, China}
\affiliation{School of Astronomy and Space Science, University of Science and Technology of China, Hefei 230026, China}

\begin{abstract}
Several Pulsar Timing Array (PTA) collaborations have recently provided strong evidence for a nHz Stochastic Gravitational-Wave Background (SGWB).
Here we investigate the implications of a first-order phase transition occurring within the early Universe's dark quantum chromodynamics (dQCD) epoch,
specifically within the framework of the mirror twin Higgs dark sector model.
Our analysis indicates a distinguishable SGWB signal originating from this phase transition, which can explain the measurements obtained by PTAs.
Remarkably, a significant portion of the parameter space
for the SGWB signal also effectively resolves the existing tensions in both the $H_0$ and $S_8$ measurements in Cosmology.
This intriguing correlation suggests a possible common origin of these three phenomena
for $0.2 < \Delta N_{\rm eff} < 0.5$, where the mirror dark matter component constitutes less than $30\%$ of the total dark matter abundance.
Next-generation CMB experiments such as CMB-S4 can test this parameter region.\\
\par\textbf{Keywords: }Mirror Twin Higgs, Dark Matter, Phase Transition, Gravitational Wave, Pulsar Timing Array

\end{abstract}

\keywords{Mirror Twin Higgs, Dark Matter, Phase Transition, Gravitational Wave, Pulsar Timing Array}

\maketitle

\section{Introduction}
Thanks to the worldwide joint efforts, the pulsar timing array (PTA) experiments, such as the North American Nanohertz Observatory for Gravitational Waves (NANOGrav)~\cite{NANOGrav:2020bcs}, Parkes Pulsar Timing Array (PPTA)~\cite{Kerr:2020qdo,Goncharov:2021oub} and European Pulsar Timing Array (EPTA) \cite{Chen:2021rqp}, had found evidence for a common spectrum noise at frequencies around $10^{-8}$ Hz. The further analysis of the PTA data (including  CPTA \cite{CPTA}, EPTA \cite{EPTA}, NANOGrav \cite{NANOGrav}, PPTA \cite{PPTA})
successfully reveals the Hellings–Downs
spatial correlations and is strongly in favor of the stochastic gravitational-wave background (SGWB) nature.
Although the frequency of the PTA SGWB signal falls within the Standard Model Quantum Chromodynamics (QCD) scale,
it is widely speculated that the QCD phase transition (PT) is plausibly a crossover rather than a first-order transition. Consequently, it is not expected to generate a detectable SGWB signal~\cite{Aoki:2006we, Bhattacharya:2014ara, Kajantie:1996mn}.
Alternatively, the mergers of the supermassive black hole binaries (SMBHB) are natural astrophysical sources of SGWB.
Given the current limitations in detection capabilities, identifying individual events involving SMBHB remains a challenge. We also face difficulties such as the ``last parsec problem" and insufficient local density in explaining the origin of the SGWB~\cite{Casey-Clyde:2021xro, Kelley:2016gse, Kelley:2017lek,Oikonomou:2023qfz}.
Anyhow, we expect to better understand the problem with future rich observational data and theoretical developments.
Furthermore, the cosmological origin of SGWB is being extensively discussed~\cite{Ashoorioon:2022raz,Athron:2023xlk,Arunasalam:2017ajm,Kobakhidze:2017mru,Oikonomou:2023qfz,Vagnozzi:2023lwo,Oikonomou:2023bli,Addazi:2023jvg,Wang:2023len},
and may shed light on new physics beyond the Standard Model (SM).

However, with rich observational data and more theoretical developments,
we expect to obtain a better understanding of the origin of the SGWB.
Furthermore, the cosmological origin of SGWB is being extensively discussed,
offering valuable insights into potential new physics beyond the SM.

One well-motivated model capable of generating a first-order PT (and hence the nHz SGWB) is the mirror twin Higgs (MTH). The MTH offers a compelling framework for addressing the Higgs hierarchy problem by introducing a softly broken mirror symmetry into the SM~\cite{Chacko_2006,Chacko2_2006}.
This symmetry implies corresponding counterparts for the particles in the SM,
albeit with larger masses due to higher vacuum expectation values (VEV). Moreover, the mirror quark twin sector, which features an twin 
$Z_2$ symmetry, interacts with a novel $\rm{SU}_{\rm twin}(3)$ gauge group, leading to predictions of a strong first-order PT capable of generating a detectable SGWB~\cite{Schwaller:2015tja, Barbieri:2016zxn, Fujikura:2018duw,Badziak:2022ltm}.

Intriguingly, MTH cosmology presents an viable scenario
for resolving tensions of both $H_0$ and $S_8$~\cite{Zu:2023rmc, Bansal:2021dfh}.
Observations, such as the Planck Cosmic Microwave Background (CMB),
suggest that the mirror sector possesses a significantly lower initial temperature.
The temperature in the twin sector impacts not only the additional radiation components that arise from twin photons but also the timing of the dark Quantum Chromodynamics (dQCD) PT.
Moreover, a larger twin VEV predicts an increased confinement scale for twin QCD ($\Lambda_{d\rm{QCD}}$).
As a result, the temperature and VEV parameters associated with broken symmetry contribute to an earlier onset of the dark QCD phase transition,
coinciding with the phase transition in the SM.
Consequently, the SGWB offers a unique opportunity to test the MTH cosmology and complement cosmological observations in the electromagnetic spectrum.

\section{SGWB in the mirror twin Higgs model}
Compared with the VEVs, the $u$ and $d$ quarks are nearly massless, while the $s$ quark has a mass of the order of the QCD confinement scale.
The Lattice simulations, including these physical quark masses, have demonstrated that the QCD phase transition is a crossover rather than a first-order transition~\cite{Aoki:2006we, Bhattacharya:2014ara}. However, it is important to note that this result is specifically based on the physical quark masses in the SM. If the dark sector possesses a ${\rm{SU}}(N_d)$ gauge group theory whose $N_d$ corresponds to the dQCD group but with different masses for the dark quarks, it is conceivable to have a first-order phase transition (PT) and generate the stochastic gravitational waves (SGW)~\cite{Schwaller:2015tja, Garani:2021zrr, Bringmann:2023opz}.

In the case of a dark sector with an ${\rm SU}(N_d)$ gauge group theory together with a total of $N_f$ nearly massless fermions and a confinement scale $\Lambda_d$, the phase transition is the first order if either $N_f=0$ or $3 \leq N_f < 4N_d$ \cite{Schwaller:2015tja, Panero:2009tv, PhysRevD.29.338}. However, the situation becomes more intricate when considering the case of $N_f=2$ and finite $N_d$, as analytic arguments cannot be readily applied, and lattice simulations are challenging~\cite{Aoki:2006we, Bhattacharya:2014ara}. Moreover, in the context of the MTH, if the mirror symmetry remains unbroken, the case of $N_f=2$ and $N_d=3$ is analogous to the SM and still difficult to treat the problem. Therefore, in this study, we consider two cases: (i) $N_f=0$, the minimal setup assuming only the presence of twin top and bottom quarks that are necessary for addressing the hierarchy problem~\cite{Chacko_2006, Chacko2_2006}, and (ii) $N_f=4$, an extended scenario assuming two additional nearly massless flavours. \footnote{Here,  the ratio of two VEVs is assumed to be not excessively large ($<15$) to maintain the naturalness of the theory. If we do not require it to address the naturalness problem, it opens up the possibility of achieving a scenario with $N_f=0$ while maintaining an unbroken mirror symmetry.}

Previous studies have demonstrated that the twin sector should be colder than the normal sector~\cite{Bansal:2021dfh, Farina:2015uea, Barbieri:2016zxn}.
One approach to achieving this temperature disparity is introducing an asymmetric post-inflationary reheating mechanism that is more efficient in the SM sector than in the twin sector~\cite{Chacko:2016hvu, Craig:2016lyx}. In this study, we denote the temperature ratio at dQCD PT as $\epsilon = T_{\rm{twin}}/T_{\rm{SM}}$.
On the other hand, the confinement scale for twin QCD ($\Lambda_{d\rm{QCD}}$) is higher than that of the SM ($\Lambda_{\rm{QCD}}$),
even for equal gauge couplings due to the role of the higher twin VEV in the coupling running.
The confinement scales for the two sectors at the one-loop level can be expressed as~\cite{Bansal:2021dfh}
\begin{equation}
    \frac{\Lambda_{d \rm QCD}}{\Lambda_{\rm QCD}} \sim 0.68+0.41\lg\left(1.32+\frac{\hat{v}}{v}\right),
\end{equation}
where $\hat{v}$ and $v$ represent the VEV of the twin and SM Higgs.
As a result, the temperature of the dQCD PT is higher than that of the SM by a factor of $\Lambda_{d\rm{QCD}}/\Lambda_{\rm{QCD}}$.

A first-order PT would generate gravitational waves through various mechanisms such as bubble collisions~\cite{Kosowsky:1992vn, Caprini:2007xq}, sound waves~\cite{Hindmarsh:2017gnf, Ellis:2020awk}, and magnetohydrodynamical (MHD) turbulence~\cite{Kosowsky:2001xp, Dolgov:2002ra, Kahniashvili:2008pf}.
Following the approaches of~\cite{NANOGrav:2021flc, Bringmann:2023opz}, we consider the contributions from bubble collisions (BC) and sound waves (SW),
as the MHD contribution is sub-leading compared to the sound-wave contribution and subjects to large uncertainties~\cite{Caprini:2015zlo, RoperPol:2019wvy}.

As revealed in the simulations, the present-day differential energy density in the gravitational wave background (GWB), as a function of the gravitational wave frequency, can be described as~\cite{Caprini:2015zlo, Binetruy:2012ze, Cutting:2020nla}
\begin{equation}
\begin{aligned}
    &h^2\Omega_{\rm{GW}}(f)=7.69\times10^{-5} g^{-\frac{4}{3}}_{s*}g_{\rho *} \left(\frac{\kappa_{\phi(sw)} \alpha_*}{1+\alpha _*}\right)^2 \\
    &\begin{cases}
\frac{0.48v^3_w}{1+5.3v^2_w+5v^4_w}
\frac{H_*^2}{\beta^2}
S_b\left(\frac{f}{f_{p,b}}\right),  \quad  {\rm for~bubble~collision},
 \\
0.513v_w
\frac{H_*}{\beta} S_{sw}\left(\frac{f}{f_{p,sw}}\right) \Gamma(\tau_{sw}),
\quad  {\rm for~sound~wave},
\end{cases}
\end{aligned}
\end{equation}
where $v_w$ is the bubble wall velocity, $\kappa_{\phi(sw)}$ is the energy conversion efficiency, and in our calculation, $v_w=1$ and $\kappa=1$ are assumed for large energy conversion efficiency from vacuum energy to bulk fluid motion~\cite{Bringmann:2023opz}.
The $g_{s*}~(g_{\rho *})$ is the effective number of relativistic degrees of freedom contributing to the total entropy (energy) at the time of
production, $\alpha_*$ is the strength of the phase transition, the ratio of the vacuum and relativistic energy density at the time of phase transition and $\beta/H_{*}$ is the bubble nucleation rate in units of the phase transition Hubble rate.
The function $S_{b(sw)}$ describes the spectral shape of the BC (SW), and $f_{p,b(sw)}$ represents the peak frequencies. In this study, we adopt the benchmark point $(a, b, c) = (0.7, 2.3, 1)$ in line with the numerical simulation findings~\cite{NANOGrav:2021flc}.
The suppression factor $\Gamma(\tau_{sw})=1-(1+2\tau_{sw}H_*)^{-1/2}$ owes to the finite lifetime of the sound waves~\cite{Guo:2020grp,Ellis:2020awk}.
The  functional form found in the lattice simulation reads~\cite{Hindmarsh:2017gnf,Cutting:2020nla}
\begin{equation}
\frac{f_{p,b(sw)}}{0.113~{\rm{nHz}}} =
\left(\frac{f_*}{\beta}\right)_{b(sw)} \left(\frac{\beta}{H_*}\right) \left(\frac{T_*}{\rm{MeV}}\right)
\sqrt{\frac{g_{\rho *}}{10}} \left(\frac{10}{g_{s *}}\right)^{\frac{1}{3}},
\end{equation}
where
\begin{equation}
\begin{aligned}
   \left(\frac{f_*}{\beta}\right)_{b}&=\frac{0.35}{1+0.07v_w+0.69v^4_w},
        \\
   \left(\frac{f_*}{\beta}\right)_{sw}&=\frac{0.536}{v_w},
\end{aligned}
\end{equation}
and
\begin{equation}
\begin{aligned}
S_b(x)&=\frac{(a+b)^c}{(bx^{-a/c}+ax^{b/c})^c}, \\
S_{sw}(x)&=x^3\left(\frac{7}{4+3x^2}\right)^{7/2}.
\end{aligned}
\end{equation}
The temperature of the normal sector at the time of the PT is denoted as $T_*$. In the context of the MTH model, we have $\epsilon T_* = T_{d\rm{QCD}} = \frac{\Lambda_{d\rm{QCD}}}{\Lambda_{\rm{QCD}}} \times T_{\rm{cr}}$, where $T_{\rm{cr}}$ is the critical phase temperature ($T_{\rm{cr}} \sim 235 \rm{MeV}$ for case (i) and $T_{\rm{cr}} \sim 135 \rm{MeV}$ for case (ii))~\cite{Braun:2006jd}. Additionally, the twin sector contributes to $g_*$ through its additional relativistic components. We have $g_{s *} = g_{\rm{SM}} + g_{\rm{twin}}\epsilon^3$ and $g_{\rho_*} = g_{\rm{SM}} + g_{\rm{twin}}\epsilon^4$, where $g_{\rm{SM}} = 47.75$ and $g_{\rm{twin}} = 47.75 - 36 \times \frac{7}{8}$ for the minimal scenario, and $g_{\rm{twin}} = 47.75 + 24 \times \frac{7}{8}$ for the extended scenario. While our results indicate that the SGWB spectra for both scenarios are nearly indistinguishable within our interesting parameter space but with a shift of the parameter $\sigma$, in the following analysis, we focus on presenting the results for the extended case (ii).

Analytical calculation of the dynamics of the PT
is possible.
However, now we are dealing with a strongly coupled regime where such analytical methods are not applicable.
Therefore, we consider $\beta/H_*$, $\epsilon$, $\hat{v}/v$, and $\alpha_*$ as the four free input parameters
and assign reasonable values
found in weakly coupled models and simulations~\cite{Schwaller:2015tja, Bringmann:2023opz,Freese:2022qrl}.
It is important to note that
these parameters are correlated with each other in principle.
Our treatment (the four parameters are uncorrelated) would degenerate to a generic dQCD PT process in some cases,
including not only the MTH model but also other dQCD PT models.
Nevertheless,
more careful treatment of this correlation can improve the current investigation
and we plan to revisit this issue in the future.

\begin{figure}[t!]
    \centering
    \includegraphics[scale = 0.36]{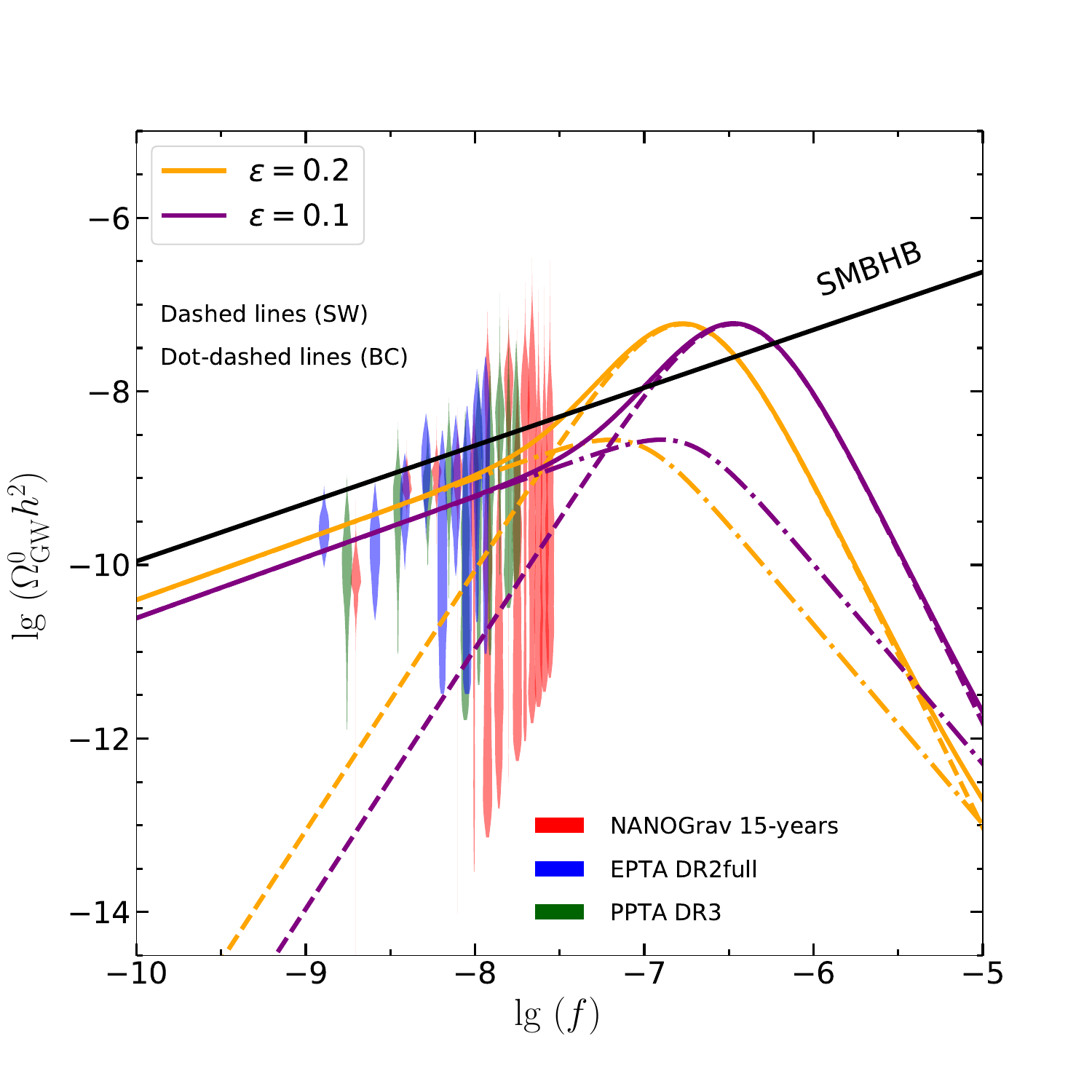}
    \caption{The predicted SGWB spectrum from the MTH dQCD PT. The dash-dotted lines represent the BC contribution and the dash lines for the SW.
    The solid lines are the total SGWB spectra,
    while the orange and purple lines represent the $\epsilon$ equal to 0.2 and 0.1, respectively.
    The red, blue, and dark green regions are the tentative SGWB signals from NANOGrav~\cite{NANOGrav}, EPTA~\cite{EPTA}, and PPTA~\cite{PPTA}.
    The SGWB from SMBHBs is also shown as a black line.
    }
    \label{Fig:MTH_gw}
\end{figure}

In Fig.~\ref{Fig:MTH_gw}, we display the SGWB spectra from the MTH dQCD PT by fixing $\beta/H_* = 3$, $\hat{v}/v = 5$, and $\alpha_* = 0.2$,
with $\epsilon=0.2$ (orange lines) and $\epsilon=0.1$ (purple lines).
An SMBHB contribution (black line) is also shown with
spectrum power $\gamma=4/3$ and amplitude $A=2\times 10^{-15}$.

A lower value of $\epsilon$ corresponds to the hotter normal sector and thus a higher peak frequency in the spectrum.
The BC component dominates the lower frequency region,
while the SW contributions are more important at the higher frequency region where
the uncertainties of PTA data are larger.

\section{The fit to the SGWB data}
 In PTA experiments, the energy density of GW is commonly related to the GW characteristic strain spectrum through the equation~\cite{Moore:2014lga,NANOGrav:2021flc}
\begin{equation}
    \Omega_{\rm{GW}}(f)=\frac{2\pi^2}{3H_0^2} f^2 h_c^2(f),
\end{equation}
where $H_0$ is the Hubble constant and $h_c(f)$ is a power-law function for the GWB characteristic strain spectrum
\begin{equation}
    h_c(f)=A_{\rm{GWB}} \left(\frac{f}{3.17\times10^{-8}~\rm{Hz}}\right)^\gamma,
\end{equation}
where $A_{\rm{GWB}}$ and $\gamma$ are the amplitude and power-law exponent.
All of the PTAs collaboration~\cite{NANOGrav,PPTA,EPTA} also adopted this power-law form to fit the 15-year dataset.
We use them to constrain the MTH model parameters.

\begin{figure}[t!]
    \centering
    \includegraphics[scale = 0.45]{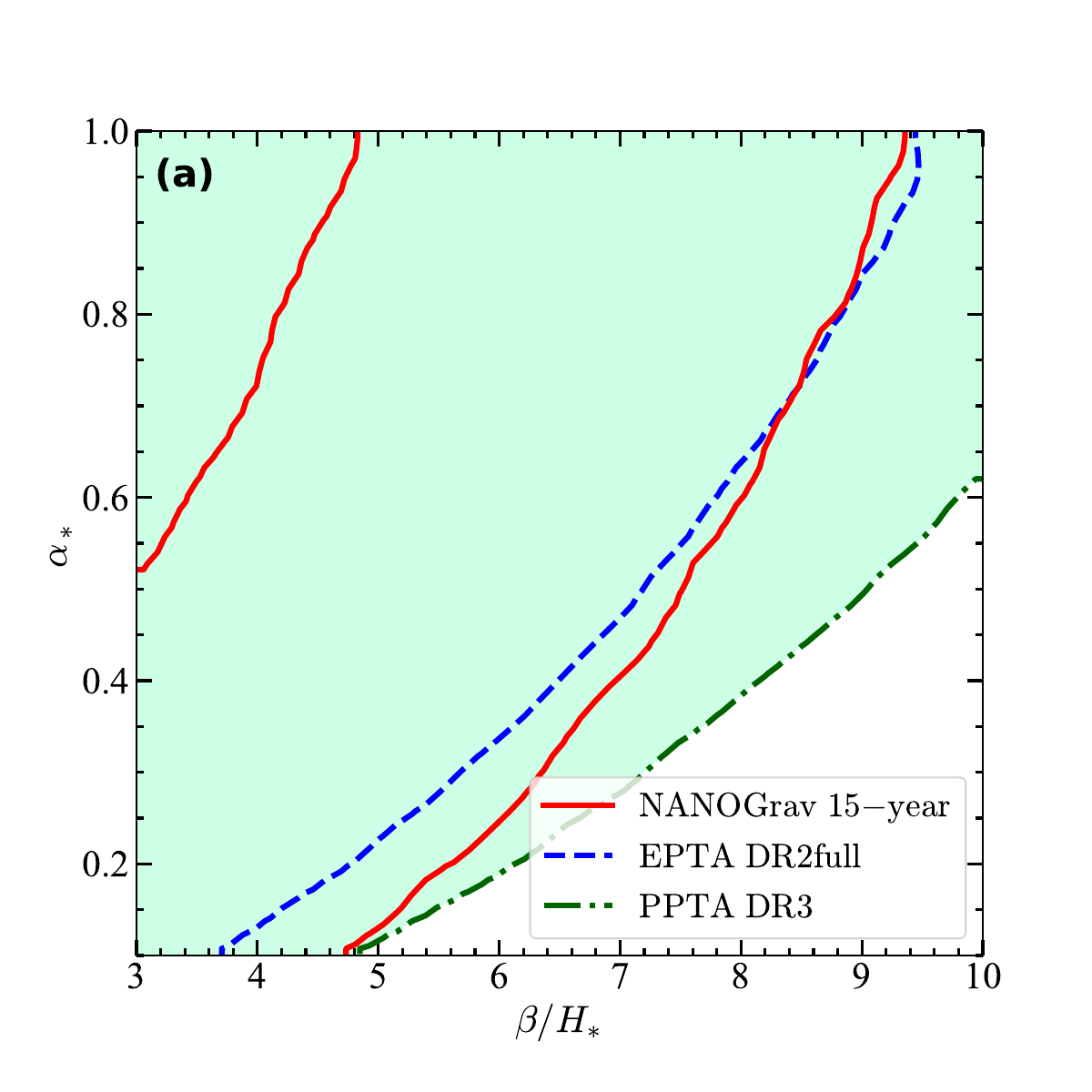}
    \includegraphics[scale = 0.45]{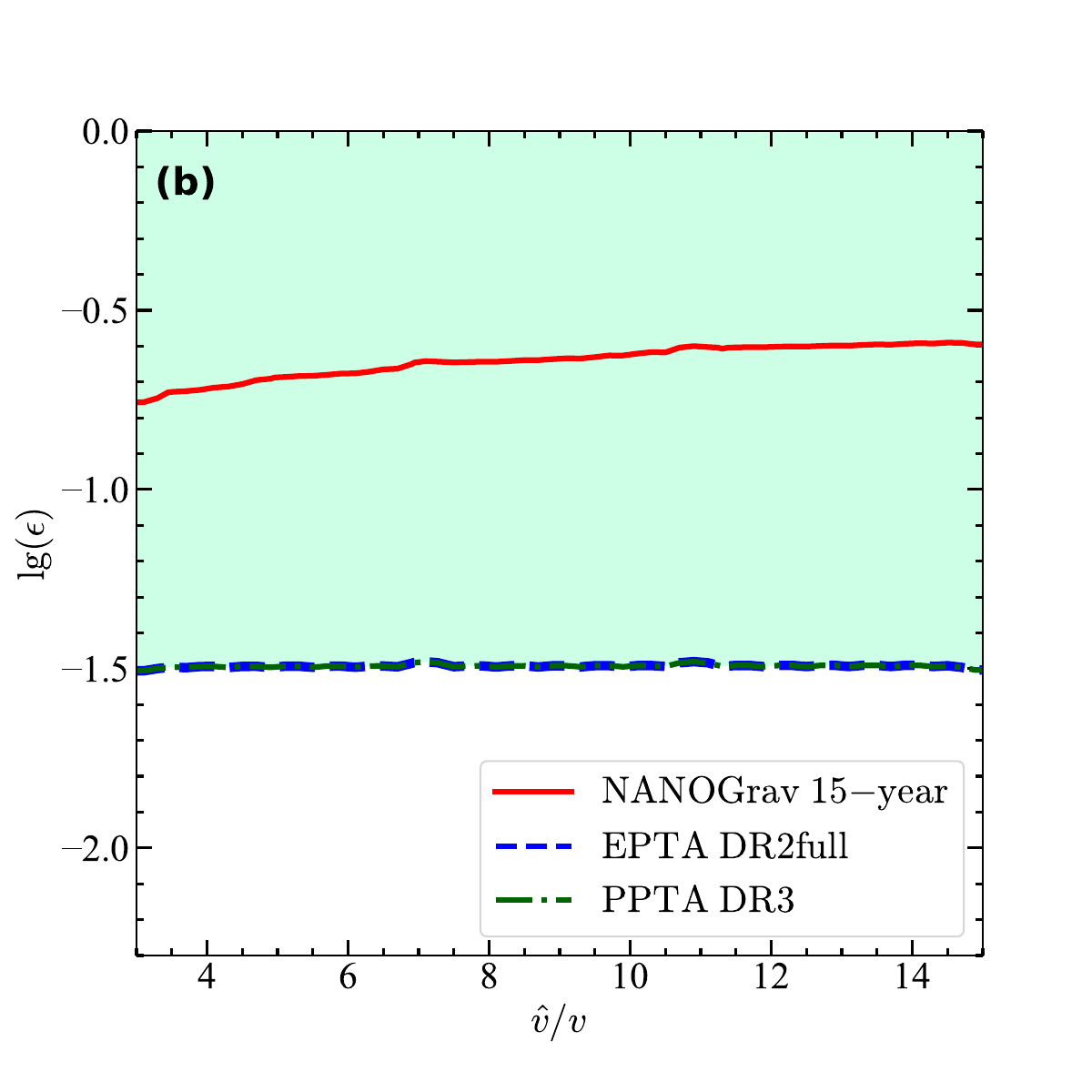}
    \caption{Allowed parameter space ($2\sigma$) accounts for the SGWB signals. The panel (a) illustrates the variables of $\beta/H_*$ and $\alpha_*$, whereas panel (b) displays $\hat{v}/v$ and $\epsilon$. The green regions are supported by NANOGrav (red), EPTA (blue) and PPTA (dark green) datasets. }
    \label{Fig:MTH_fitting}
\end{figure}

Our  $2\sigma$ contours favored by NANOGrav 15-year, PPTA and EPTA data are given in Fig.~\ref{Fig:MTH_fitting}.
Here we restricts the parameter $\beta/H_*>3$ to ensure the bubble percolation~\cite{Freese:2022qrl}. The upper plot clearly demonstrates that a strong phase transition (characterized by a large $\alpha$) and a slow transition rate (indicated by a small  $\beta$) are favored by all experiments. Meanwhile, the bottom plot shows the MTH parameter space ($\epsilon$ and $\hat{v}/v$).
Since the signal frequency (sensitive to the dark sector temperature) detected by NANOGrav is lower than the EPTA and PPTA~\cite{NANOGrav,EPTA,PPTA},
the preferred value of $\epsilon$ by NANOGrav is larger than the other experiments.
Note that to account for the SGWB signal in all three experiments, a non-zero value of $\epsilon$ is required.
However, a sufficiently small value of $\epsilon$ can mimic the standard $\Lambda\rm{CDM}$ cosmology~\cite{Bansal:2021dfh, Zu:2023rmc}.
This indicates that the SGWB signal hints at physics beyond the standard cosmological model.
Within the MTH framework, the twin sector cannot be excessively cold because it influences the timing of the dQCD PT.
Consequently, this has implications for cosmological observations such as the CMB and weak lensing.
The impacts of these factors on cosmological data will be discussed in the following section.
Additionally, Fig.~\ref{Fig:MTH_fitting} demonstrates that the effect of $\hat{v}/v$ is relatively small, and the distributions of $\beta/H_*$ and $\alpha_*$ agree well with a previous model-independent work based on the 12.5-year NANOGrav dataset~\cite{NANOGrav:2021flc}.

\begin{figure}[h]
    \centering
    \includegraphics[scale =0.45]{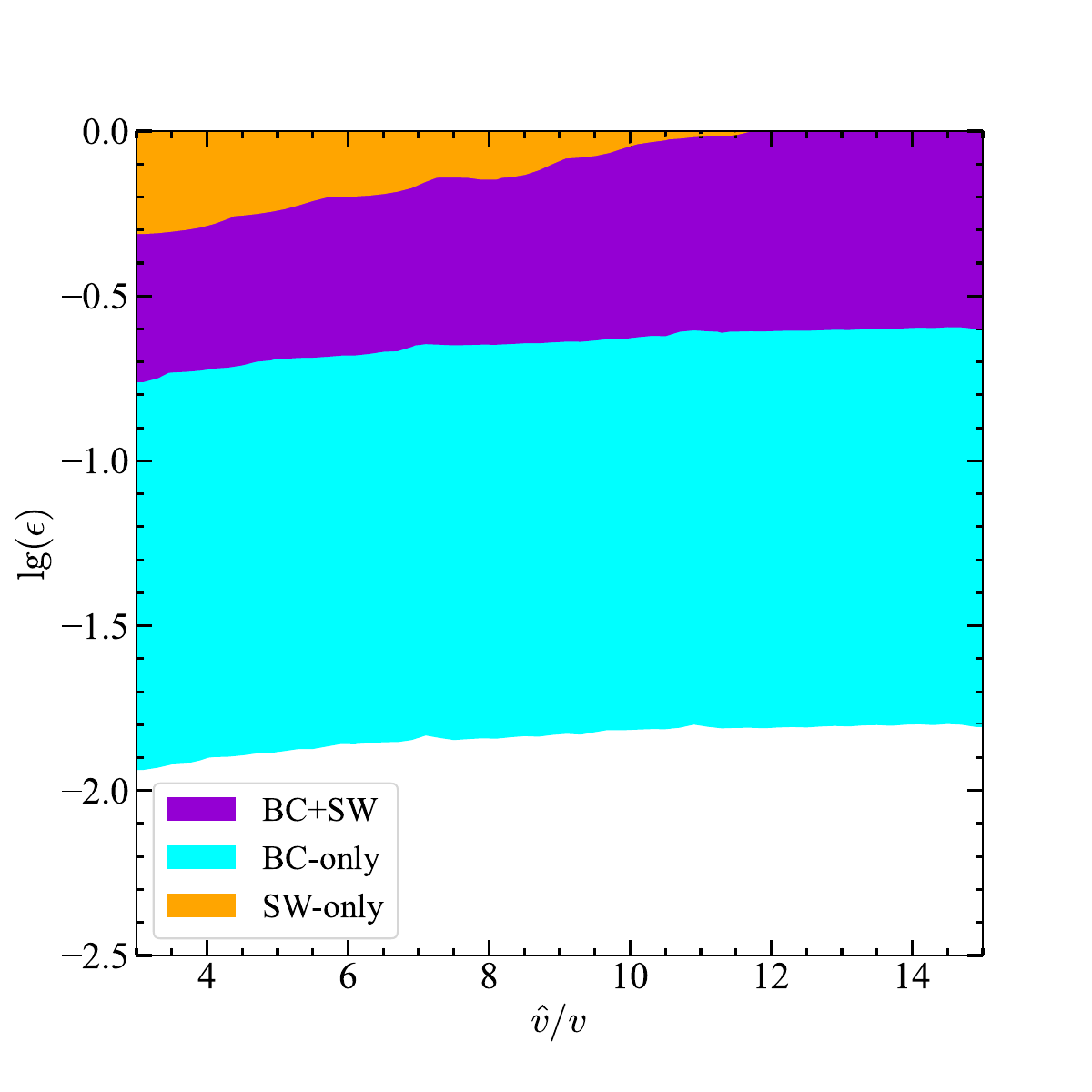}
    \caption{The 2$\sigma$ allowed parameter space favored by NANOGrav datasets for the SW only (orange), BC only (cyan), and the SW+bubble collision contribution (purple).
    }
    \label{MTH_sw}
\end{figure}

In general, both the sound waves (SW) and bubble collisions (BC) components contribute to the SGWB.
While depending on the interaction between bubble walls and the surrounding plasma,
SW-only (BC-only) contribution is also expected if majority (minority) of the energy or are transferred to the plasma.
Here, we also show the SW-only and BC-only fitting results in Fig.~\ref{MTH_sw}.
We choose the numerical BC spectrum parameters as $(a,b,c)=(2,2.4,1.7)$ for BC.
The expected dark sector temperature is cooler (hotter) for BC (SW) since the SW peak frequency is higher than the BC.

The SGWB is not the only cosmological anomaly that has emerged. Currently, there are at least two other widely discussed cosmological tensions. The first is the $H_0$ tension, which pertains to the discrepancy between the measured value of the Hubble constant ($H_0$) and the value predicted by the $\Lambda\rm{CDM}$ model favored by the Planck CMB observations. The observed $H_0$ value is in tension with the local observations, indicating a discrepancy $\sim 5\sigma$~\cite{DiValentino:2021izs, Riess_2021}.
The second tension is related to $S_8$, a parameter characterizing the amplitude of matter fluctuations on large scales. The predicted value of $S_8$ based on the $\Lambda\rm{CDM}$ model is in 2-3 $\sigma$ tension with observations from weak lensing surveys, such as the Dark Energy Survey (DES)~\cite{MacCrann:2014wfa, DES:2021wwk}.

These cosmological and astrophysical observations, including the SGWB signals, collectively indicate the presence of new physics beyond the standard cosmological model, $\Lambda\rm{CDM}$.
In our previous work~\cite{Zu:2023rmc}, we have investigated the MTH cosmology by combining weak lensing data from the Dark Energy Survey (DES), Planck CMB data, and baryon acoustic oscillation (BAO) observations.
In this work, we investigate a fundamental process arising from the interaction of DM and dark radiation, thus the core conclusions of our cosmological study should still remain reasonable.
The analysis in Ref.~\cite{Zu:2023rmc} reveals that the MTH model can potentially alleviate the $H_0$ and $S_8$ tensions.  Moreover, as shown in Figure~\ref{Fig:MTH_fitting}, the MTH model can also explain the SGWB signals. Therefore, it is highly appealing to explore the possibility of a parameter space within the MTH model that can simultaneously address all of these cosmological anomalies.

We can express the effective number of extra neutrinos as $\Delta N_{\rm eff} = 4.4 \left(\frac{\hat{T}_{\rm rec}}{T_{\rm rec}}\right)^4$,
following the notation of~\cite{Bansal:2021dfh}.
In this work, we assume that the twin photon is massless and ignore the twin neutrinos.
The twin and normal sectors have different temperatures during the recombination epoch, denoted by $\hat{T}_{\rm rec}$ and $T_{\rm rec}$, respectively.
Assuming the reheating is instantaneous\footnote{It is important to mention that the actual PT occurs almost instantaneously,
as its duration is only a small fraction of a Hubble time ($\beta/H_*>3$)~\cite{Freese:2022qrl}.}, the temperature ratio at the recombination epoch is given by~\cite{Bringmann:2023opz, Bai:2021ibt}
\begin{equation}
    \left(\frac{\hat{T}_{\rm{rec}}}{T_{\rm{rec}}}\right)^4=
    \left[\alpha_*+\epsilon^4(1+\alpha_*)\frac{g_{\rm twin}}{g_{\rm{SM}}}\right]\frac{g_{\rm{SM}}}{g_{\rm twin}}.
\end{equation}
When $\alpha_*=1$ and $\epsilon=0.1$,
the twin sector temperature reaches its maximum reheating temperature, which is nine times higher than the temperature prior to reheating.
We can see that $\Delta N_{\rm eff}$ depends on $\epsilon^4$ and the strength of the phase transition $\alpha_*$.

\begin{figure}[t!]
    \centering
    \includegraphics[scale = 0.45]{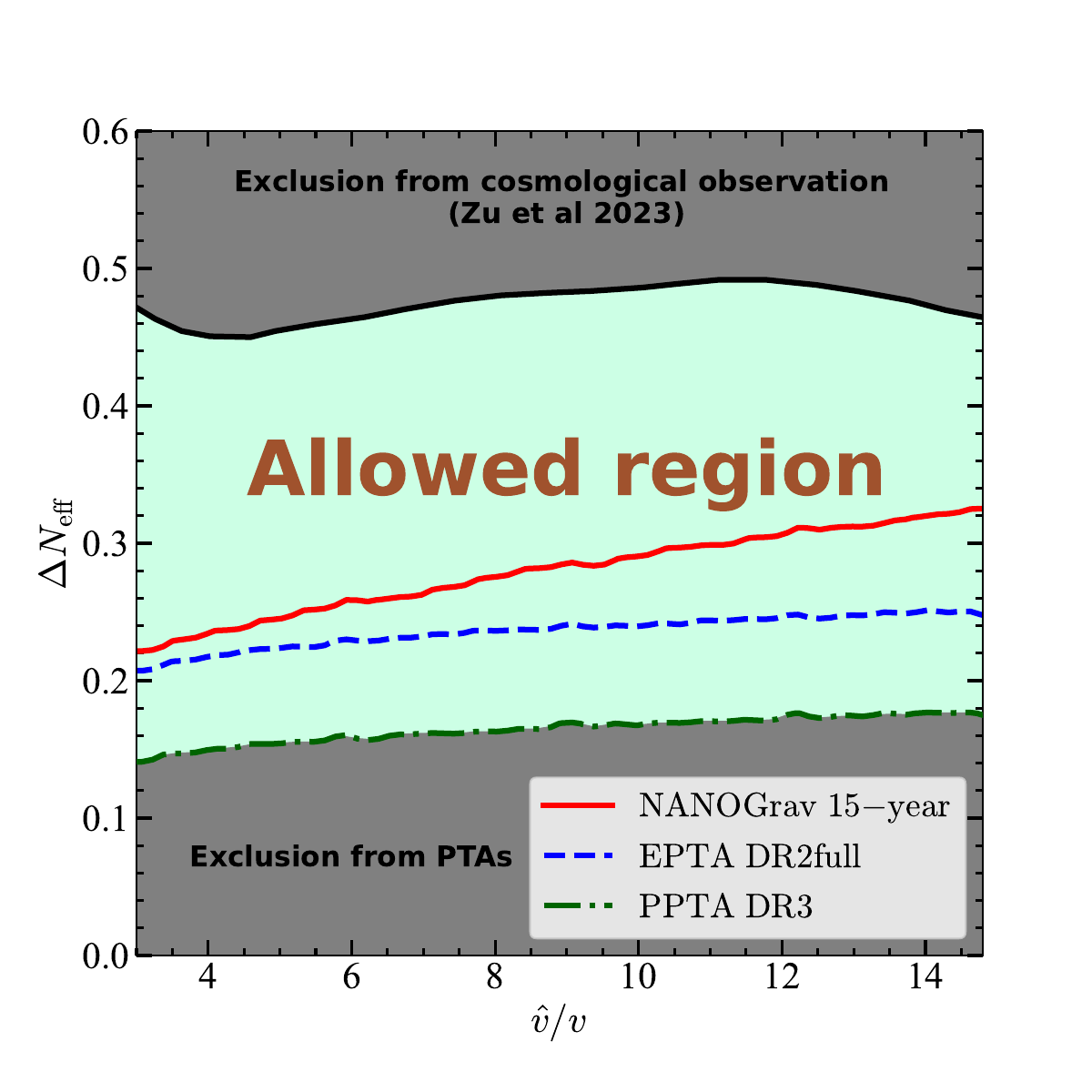}
    \caption{The correlation between the MTH physical parameters: derived $N_{\rm{eff}}$ and $\hat{v}/v$. 2$\sigma$ regions have shown for three experiments. The solid black lines are the $95\%$ upper limits from the fitting results of cosmological observations~\cite{Zu:2023rmc}.
    }
    \label{Fig:MTH_Neff}
\end{figure}

By performing the SGWB analysis, we obtain the favored parameter space of $\Delta N_{\rm eff}$.
The results are shown in Figure~\ref{Fig:MTH_Neff}.
Both $\epsilon$ and $\Delta N_{\rm eff}$ cannot vanish to generate the SGWB, namely $\epsilon>0.2$ and $\Delta N_{\rm eff} > 0.2$ in $2\sigma$ region.
Nevertheless, based on the cosmological data (Planck CMB, DES 3-year cosmic shear, BAO, and SH0ES data),
$\Delta N_{\rm eff}$ has an upper limit of $\sim 0.5$~\cite{Zu:2023rmc}.
Such the combined cosmological likelihood can probe
a slightly higher $\Delta N_{\rm eff}$ than the one obtained from the only CMB measurements ($\Delta N_{\rm eff} \leq 0.3$).
Given two constraints from SGWB and cosmological data, we find a finite parameter space region $0.2 < \Delta N_{\rm eff} < 0.5$, as shown in the bottom of Fig.~\ref{Fig:MTH_Neff}.
Importantly, this region can also explain the SGWB signal and simultaneously alleviate the $H_0$ and $S_8$ tensions.
Therefore, we would like to emphasize that the MTH cosmology presents an appealing and distinctive parameter space that addresses all of the anomalies beyond the $\Lambda\rm{CDM}$ model. Moreover, the upcoming generation of CMB observations, such as CMB-S4, holds the potential to detect $\Delta N_{\rm eff}$ with a sensitivity of $\Delta N_{\rm eff} < 0.06$ at a confidence level of $95\%$~\cite{CMB-S4:2022ght}. The more precise cosmological data in the near future will stringently test the above parameter range.

Unlike Ref.~\cite{Bringmann:2023opz}, our study exclusively examines the MTH model.
Our results differ from Ref.~\cite{Bringmann:2023opz} in two key ways.
Firstly, we impose $\beta/H_*>3$ to ensure percolation~\cite{Freese:2022qrl},
as opposed to the more conservative criterion of $\beta/H_*>10$.
Secondly, we attribute the contribution of $\Delta N_{\rm eff}$ to the twin photons after twin recombination.
This new contribution is distinct from the conventional QCD PT.
By accounting for these factors, the MTH model can simultaneously explain the SGWB and alleviate the $H_0$ and $S_8$ tension.

\section{Conclusion}
We have investigated MTH cosmology with a dQCD PT that can generate nHz gravitational wave radiation.
While the MTH was initially proposed to address the Higgs hierarchy problem, it can reasonably account for the SGWB recently detected by several PTA experiments.
Moreover, the allowed parameter space can effectively alleviate the well-known $H_0$ and $S_8$ tensions in Cosmology.
Remarkably, these cosmological observations are correlated to the Higgs hierarchy problem and occur at different stages of cosmic evolution.
By combining our previous cosmological fitting results~\cite{Zu:2023rmc} with the current SGWB fit result,
we identify a surviving parameter space $0.2 < \Delta N_{\rm eff} < 0.5$ where mirror DM composition is less than $30\%$ of total dark abundance,
which agrees with all current cosmological observations. Future experiments in multi-frequency gravitational wave detection,
powerful galaxy surveys~\cite{Gong:2019yxt,EUCLID:2011zbd}, and next-generation CMB-S4~\cite{CMB-S4:2022ght}
can completely probe this parameter space.\\

{\bf Conflict of interest}\\
The authors declare that they have no conflict of interest.\\

{\bf Acknowledgments}\\
We thank the referees for helpful suggestions to further improve this manuscript.
This work was supported by the National Key Research and Development Program of China (2022YFF0503304 and  2022YFF0503301), the Natural Science Foundation of China (11921003 and 12003069), the New Cornerstone Science Foundation through the XPLORER PRIZE, the Chinese Academy of Sciences, and the Entrepreneurship and Innovation Program of Jiangsu Province.\\

{\bf Author contributions}\\
{Yi-Zhong Fan, Yue-Lin Sming Tsai, and Lei Zu conceived the idea.
Lei Zu, Chi Zhang, Yu-Chao Gu, and Yao-Yu Li conducted the numerical calculations.
All authors discussed the results and wrote the manuscript.}
~\\

\bibliographystyle{unsrt}

\end{document}